\documentclass[prb,aps,floatfix,superscriptaddress,twocolumn]{revtex4}
\usepackage{amsmath}
\usepackage[final]{graphicx}
\usepackage{dcolumn}
\usepackage{bm}
\begin{document}

\title{\textit{Ab initio} study of the modification of elastic properties of $\alpha$ iron by hydrostatic strain and by hydrogen interstitials}

\author{D.~Psiachos}
\affiliation{ICAMS, Ruhr-Universit{\"a}t Bochum, Bochum, Germany}

\author{T.~Hammerschmidt}
\affiliation{ICAMS, Ruhr-Universit{\"a}t Bochum, Bochum, Germany}

\author{R.~Drautz}
\affiliation{ICAMS, Ruhr-Universit{\"a}t Bochum, Bochum, Germany}

\begin{abstract}
The effect of hydrostatic strain and of interstitial hydrogen on the elastic properties of $\alpha$-iron is investigated using 
\textit{ab initio} density-functional theory calculations. We find that the cubic elastic constants and the
polycrystalline elastic moduli to a good approximation decrease linearly with increasing hydrogen concentration. This net strength reduction
can be partitioned into a strengthening electronic effect which is overcome by a softening volumetric effect. The calculated 
hydrogen-dependent elastic constants are used to determine the polycrystalline elastic moduli and anisotropic
elastic shear moduli. For the key slip planes in $\alpha$-iron,
$\left[1\overline{1}0\right]$ and $\left[11\overline{2}\right]$, we find a shear modulus
reduction of approximately 1.6\% per at.\% H.
\end{abstract}

\maketitle

\section{Introduction}

Hydrogen degrades the performance of many alloys and steels by lowering the failure stress, leading to fracture at unpredictable loading
conditions~\cite{Oriani,Hirth}. Several mechanisms have been proposed to explain the H-embrittlement of iron, the main ones being the
hydrogen-enhanced decohesion (HEDE) mechanism~\cite{Troiano,Oriani2}, H-vacancy effects~\cite{DawBaskes,Ohno,TateyamaPRB}, and
hydrogen-enhanced localised plasticity (HELP)~\cite{Beachem,BirnbaumMatSciEng,Robertson}.

In the HEDE mechanism, H weakens the cohesive bonds between the metal atoms, leading to failure at interfaces where H tends 
to concentrate in, such as around the tensile strain field of a crack
opening~\cite{Troiano,Oriani2}.
Vacancies containing H can order themselves along critical slip directions, leading to fracture~\cite{Ohno,TateyamaPRB}.
Within the HELP mechanism, the onset of plasticity with loading occurs at a lower stress as a result of the H-shielding of repulsive
interactions between dislocations~\cite{BirnbaumMatSciEng}.
The increased H concentration at dislocations~\cite{Cottrell} effectively reduces the dislocation-dislocation spacings. The 
resulting phenomenon of dislocation
coalescence, and ultimately, crack advancement at reduced loads, in the presence of H, may be related to the experimentally-observed 
increase of dislocation mobility caused by H~\cite{Robertson}.
The importance of increased H-concentration near dislocations and other low-energy trap sites such as interstitial sites
was also indicated by a recent experimental
study of intergranular failure in steel~\cite{Novak}. In that study, the fracture mode changed from ductile to brittle, as the amount of H
located at these low-energy trap sites increased.

One of the central challenges in avoiding hydrogen embrittlement is the interpretation of the experimentally observed
effective behaviour~\cite{Gerberich,Novak,Barnoush,Kirchheim} that arises from the interplay of H solubility and diffusivity with
cohesive/elastic properties, vacancies, dislocations, and other defects.
The advantage of theroretical approaches is that the effects can be investigated independently, in contrast to experiment. Combining
 highly-accurate electronic structure calculations\cite{JiangCarter,Sanchez,SanchezMD} with other methods enables access to
 extended time and length scales which are necessary for describing e.g. long-range strain fields or small H concentrations~\cite{Taketomi2},
kinetic effects~\cite{Rama,BecquartHeW} and for predicting continuum-scale properties~\cite{ClouetActaMat,Gavriljuk}.

The goal of this study is to establish a link between highly-accurate \textit{ab-initio} calculations and continuum elasticity-theory in order
to explain the experimentally-observed influence of hydrogen on the elastic properties of iron (\textit{e.g.} Ref.~\onlinecite{Zhang}) and
steel (\textit{e.g.} Ref.~\onlinecite{Ortiz}).
To this end we use density-functional theory to calculate the elastic constants of $\alpha$-Fe as a function of
hydrostatic stress and different concentrations of interstitial hydrogen.

In Sec.~\ref{sec:DFT} we describe the details of our DFT calculations used to calculate the elastic constants of pure Fe
(Sec.~\ref{sec:elastic}) and the modification of the elastic constants by H (Sec.~\ref{sec:Hconcsection}). We utilize
the elastic constants to calculate the effect of H on the strength parameters - bulk, Young's, and shear moduli - in Sec.~\ref{sec:moduli}
and summarise our findings in Sec.~\ref{sec:conclusions}.

\section{Details of the calculations}
\label{sec:DFT}

\subsection{\textit{Ab initio} total energies}

The calculation of elastic properties presented here is based on numerical derivatives of the total energy with respect to strain.
The large unit cells required for reaching H concentrations of a few percent makes this problem just within the scope of
present DFT calculations.

Our spin-polarised first-principles density functional theory calculations were performed using the VASP~\cite{vasp1,vasp2,vasp3} 
code. We 
used the projector augmented-wave
method~\cite{paw1,paw2}, with pseudopotentials considering the 3p electrons of Fe as valence electrons. The generalised gradient
approximation (GGA) in the PW91 parametrisation~\cite{pw91} was used for the exchange-correlation functional.
In order to verify the reliability of our conclusions, we repeated some of our calculations with
the Vosko-Wilk-Nusair~\cite{vwn} (VWN) spin interpolation in PW91, 
and the Perdew-Burke-Ernzerhof~\cite{pbe} (PBE) exchange-correlation functional.
The spin-polarised GGA has proven to give reliable results for the ground state \cite{Leung} and elastic properties \cite{Cerny} for Fe.

We used supercell geometries, a plane-wave basis with a cutoff of 500~eV and a $\Gamma$-centred k-point grid equivalent
to 18$\times$18$\times$18 for the two-atom basis bcc unit cell except in the case of the 128-atom unit cell where
the Brillouin-zone sampling was equivalent to 20$\times$20$\times$20.
We found these settings to be more than adequate for capturing the equilibrium properties, but
necessary for converging the elastic constants to within less than one percent error (see Sec. \ref{sec:elasticconv}).
The ions were relaxed until the maximum force component on each ion was less than 0.01~eV/\AA~
 while the total energies were converged to within 0.01 meV. The reported lattice parameter was determined from fitting
the total energies to the Murnaghan equation of state~\cite{murn}.

Our results for the lattice parameter (2.832~\AA), bulk modulus (194~GPa), bulk-modulus pressure derivative (5.42),
 and magnetic moment (2.17~$\mu_B$) are
in good agreement with other first-principles calculations~\cite{herper,clatterbuck,nl_th3,sha2} and
 with experimental results~\cite{rayne,acet}.

\section{Elastic Constants}
\label{sec:elastic}
\subsection{General description and convergence criteria}
\label{sec:elastic1}
The total energy of a solid at zero stress and equilibrium volume $V_0$ can be expanded about small strains $\boldsymbol{\epsilon}$
\begin{equation}
E(\boldsymbol{\epsilon})=E(0)+\frac{1}{2!}V_0\sum_{ijkl}C_{ijkl}\epsilon_{ij}\epsilon_{kl}+\ldots 
\label{e_series}
\end{equation}
where the indices run from 1-3. As the strain tensors are symmetric, the 
notation $C_{ijkl}$ can be expressed in the two-index (Voigt) form $C_{ij}$
where the indices run from 1-6. In this section we restrict ourselves to linear elastic behaviour, \textit{i.e.} stress
linear in the strain. 

For the calculation of the elastic constants in this work, we enforced cubic symmetry
of the unit cells in our DFT calculations. This reflects our expectation that a random
distribution of H would cause an effectively isotropic lattice distortion. In addition,
the restriction to cubic volume elements facilitates the scale-bridging
with mesoscopic approaches, such as \textit{e.g.} finite element schemes. In Sec.~\ref{cubicapprox}
we will show that the quantitative effect of this cubic constraint is in fact negligible for the
investigated unit cells.

Within our approximation of a cubic system, the number of independent elastic constants $C_{ij}$ reduces to 
three: C$_{11}$, C$_{12}$, and C$_{44}$. 
Their numerical values can be determined by applying suitable strain tensors
and taking the second derivative of Eq.~\ref{e_series} with respect to the applied strain.
The bulk modulus $B=\frac{1}{3}\left(C_{11}+2C_{12}\right)$ and its pressure derivative $B^\prime(P)$ 
are found by applying hydrostatic strains and then 
carrying out a fit to the Murnaghan equation of state. The 
values of $C^\prime=C_{11}-C_{12}$ and $C_{44}$ are found by applying volume-conserving strains. 

In the case of $C^\prime$, an orthorhombic strain 
\begin{equation} \boldsymbol{\epsilon} = \left( \begin{array}{ccc}
\delta & 0 & 0 \\
0 & -\delta & 0 \\
0 & 0 & \delta^2/(1-\delta^2) \end{array} \right) \label{mat1} \end{equation}
is applied, giving a total energy expression
\begin{equation}
E_{ortho.}(\delta)=E(0)+C^\prime V \delta^2+O[\delta^4],
\label{e_en1}
\end{equation}
while, for $C_{44}$, the strain is monoclinic
\begin{equation} \boldsymbol{\epsilon} = \left( \begin{array}{ccc}
0 & \delta/2 & 0 \\
\delta/2 & 0 & 0 \\
0 & 0 & \delta^2/(4-\delta^2) \end{array} \right).\label{mat2}\end{equation}
with a total energy
\begin{equation}
E_{mono.}(\delta)=E(0)+\frac{1}{2}C_{44} V \delta^2+O[\delta^4].
\label{e_en2}
\end{equation}

We fitted our \textit{ab initio} total energies for systems subject to nine values of strain $\delta$ between $\pm3\%$ 
to Eqs.~\ref{e_en1} and \ref{e_en2} in order to obtain $C^\prime$ and $C_{44}$.

\subsection{Convergence tests}
\label{sec:elasticconv}
\begin{figure}[htb]
\includegraphics[width=8cm]{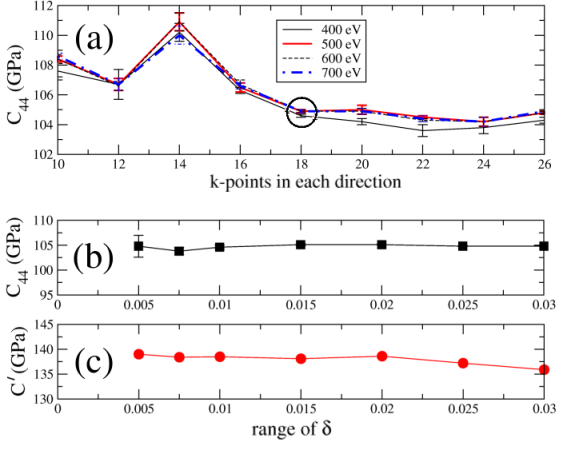}
\caption{(a) Convergence of $C_{44}$ as a function of k-point density and energy cutoff for the two-atom bcc-Fe cell
for a range of strain values $\left|\delta\right|\leq 3\%$ (see Eqs. \ref{mat1}-\ref{e_en2}). 
The error bars are deduced from the least-squares fit to Eq.~\ref{e_en2}. The circled point corresponds to the
calculation parameters used in this study. (b) $C_{44}$ and (c) $C^\prime$ obtained for different ranges of strain $\delta$.}
\label{converg}
\end{figure}

To estimate the accuracy of the elastic-constant calculations, we carried out extensive convergence
tests. The quantities varied were the k-point density, cutoff energy, and strain
$\delta$ at a lattice parameter of 2.832~\AA~. An example of these 
convergence tests is shown in Fig.~\ref{converg}a for $C_{44}$, for $\left|\delta\right|\leq$3\% (the convergence
tests for $C^\prime$ are similar). The values of cutoff energy and k-point density
of the circled point (500 eV, 18$\times$18$\times$18) were used in (b) and (c) to verify that the applied strains are
within the linear regime.

As a result of our extensive convergence tests, we observed less than a 2$\%$ variation 
in the elastic constants upon varying the range of $\delta$ between
0.5-3$\%$ change in $\delta$. For the calculations presented in the remainder of this study, we chose
 a value of 3$\%$ for $\delta$, \textit{i.e.} at the border of linear elastic behaviour,
 in order to avoid numerical 
instabilities and/or the need for very high precision total energy evaluations.

\subsection{Dependence of elastic constants on hydrostatic strain}
\label{nonlin}
In order to achieve a comprehensive description of the influence of volume-expansion due to
interstitial H, we also investigated the nonlinear elastic behaviour of the elastic constants
of pure Fe as a function of hydrostatic strain. Equivalently,
this is the effect of a volume expansion on the linear elastic constants $C_{ij}$. The modified
second-order elastic constants are given by Birch~\cite{birch} in terms of the second-order
and third-order elastic constants $C_{ij}$ and $C_{ijk}$ to linear order in the applied hydrostatic
strain. Instead of calculating the third-order elastic constants, we directly determined the variation
in the second-order elastic constants as a function of small applied hydrostatic strain.

Following Wallace [\onlinecite{Wallace}], the total energy at a volume $V$ produced by
strains applied at a reference volume $V^{ref}$ away from equilibrium,
 is modified from Eq. \ref{e_series} to include a term first-order in strain, corresponding
to hydrostatic stress $\sigma_{ij}$:
\begin{eqnarray}
E(V,\boldsymbol{\epsilon})&=&E(V^{ref},\boldsymbol{\epsilon}=0)+V^{ref}\sum_{ij}\sigma_{ij}{\epsilon}_{ij}\nonumber\\
&+&\frac{V^{ref}}{2}\sum_{ijkl}{\epsilon}_{ij}C_{ijkl}(V^{ref}){\epsilon}_{kl}.
\label{e_seriesP}
\end{eqnarray}
Formally, the strains are Lagrangian (second-order in displacement) strains evaluated with respect to $V^{ref}$ (as in Eq. 2.37 in [\onlinecite{Wallace}]) but, 
as we are only concerned with small strains, we approximate them as infinitesimal.

Because the stress is hydrostatic, it can be written as $\sigma_{ij}=-P\delta_{ij}$.
Expanding Eq.~\ref{e_seriesP} and allowing the strains represented by Eqs.~\ref{mat1} and~\ref{mat2}
to be applied with respect to $V^{ref}$, we obtain
\begin{equation}
E_{ortho.}(P,\delta)=E(P,0)+\left(C^\prime-P\right) V^{ref} \delta^2+O[\delta^4],
\label{e_en1P}
\end{equation}
\begin{equation}
E_{mono.}(P,\delta)=E(P,0)+\frac{1}{2}\left(C_{44}-\frac{P}{2}\right) V^{ref} \delta^2+O[\delta^4].
\label{e_en2P}
\end{equation}
as modifications of Eqs.~\ref{e_en1} and~\ref{e_en2}, where we have removed the $V^{ref}$ functional dependence notation
from the $C_{ij}$ for clarity.

However, the elastic constants of Eq. \ref{e_seriesP}, henceforth known as the energy-strain coefficients, 
even while modified to be valid at $V^{ref}$, are no longer equal to
the stress-strain coefficients. This is discussed in detail by Wallace \cite{Wallace} (Eq. 2.51) and Barron and Klein \cite{barron}.
 Experiments 
mostly use either ultrasonic wave-propagation~\cite{nl_exp2} or diffraction techniques~\cite{nl_exp1,nl_exp3} and obtain the pressure-varying
elastic constants from stress-strain relations. The stress-strain coefficients, which we call $\stackrel{\mathrm{\circ}}{c}_{ijkl}$
as in Ref.~\onlinecite{barron}, are related to the energy-strain coefficients $C_{ijkl}$ of Eq. \ref{e_seriesP}
by
\begin{equation}
\stackrel{\mathrm{\circ}}{c}_{ijkl}=C_{ijkl}(V^{ref})+\frac{1}{2}P\left(2\delta_{ij}\delta_{kl}-\delta_{il}\delta_{jk}-\delta_{ik}\delta_{jl}\right),
\label{otherC}
\end{equation}
or in simplified terms
\begin{eqnarray}
\stackrel{\mathrm{\circ}}{c}_{11}&=&C_{11}\nonumber\\
\stackrel{\mathrm{\circ}}{c}_{12}&=&C_{12}+P\nonumber\\
\stackrel{\mathrm{\circ}}{c}_{44}&=&C_{44}-\frac{1}{2}P.
\label{otherC2}
\end{eqnarray}
We note here that the expressions \ref{otherC}-\ref{otherC2} differ from those found in works which consider Lagrangian strains~\cite{Kimizuka,wangliyip}
but agree with others also using inifinitesimal strains~\cite{ma,marcusqiu1,marcusqiu2}.
One consequence of using the stress-strain coefficients instead of the energy-strain coefficients is that Eqs.~\ref{e_en1} and~\ref{e_en2}
retain their form, apart from the replacement of the $C^\prime$ and $C_{44}$ by $\stackrel{\mathrm{\circ}}{c}{\!}^\prime\equiv\stackrel{\mathrm{\circ}}{c}_{11}-\stackrel{\mathrm{\circ}}{c}_{12}$
and $\stackrel{\mathrm{\circ}}{c}_{44}$ respectively. 
We use the stress-strain coefficients $\stackrel{\mathrm{\circ}}{c}_{ijkl}$ in order to 
avoid ambiguities in the definition of the $C_{ijkl}$ at non-zero pressure when comparing with experiments,
or with other computational studies.

\begin{figure}[htb]
\includegraphics[width=8cm]{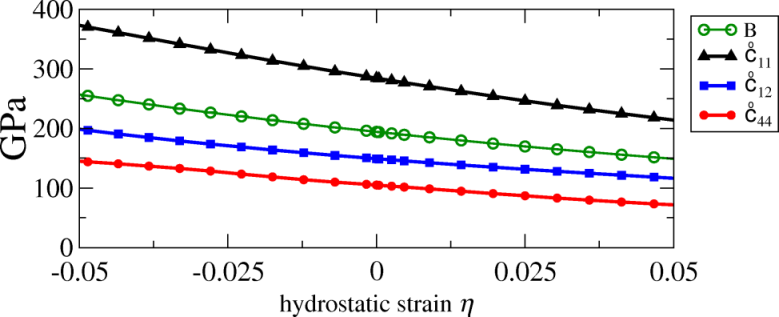}
\caption{Calculated stress-strain coefficients for pure Fe as a function of hydrostatic strain $\eta=\Delta V/V_0$
with respect to the equilibrium volume $V_0$.}
\label{linearC}
\end{figure}

At each value of hydrostatic strain $\eta$ ranging between $\pm0.05$, the value of $\delta$ 
was varied between $\pm0.03$ and a fit was made to second-order in $\delta$, as in Sec. \ref{sec:elastic1}.
These applied strain values correspond to the initial linear regime of the high-pressure transition to the body-centred-tetragonal phase \cite{ma}.
The elastic constants $\stackrel{\mathrm{\circ}}{c}{\!}^\prime$ and $\stackrel{\mathrm{\circ}}{c}_{44}$ are again extracted from the coefficients of $\delta^2$. 
However, in order to separate $\stackrel{\mathrm{\circ}}{c}_{11}$ and $\stackrel{\mathrm{\circ}}{c}_{12}$ from $\stackrel{\mathrm{\circ}}{c}{\!}^\prime$, 
we need to take into account the change of the
bulk modulus with pressure. The bulk modulus is modified by a term equal to its pressure derivative $B^\prime (P)$ 
multiplied by the pressure at the corresponding 
value of hydrostatic strain $\eta$. 
The resulting variation of the stress-strain coefficients with hydrostatic strain follows a linear trend, as can be seen
from Fig.~\ref{linearC}. 
A linear least-squares fit to the small-strain region ($\left| \eta\right|<0.02$) of these curves resulted in the following dependence
of the stress-strain coefficients on hydrostatic strain:
\begin{eqnarray}
B(\eta)&=&(194-1075\eta) \;\;\mathrm{GPa}\nonumber\\
\stackrel{\mathrm{\circ}}{c}_{11}(\eta)&=&(284-1492\eta) \;\;\mathrm{GPa}\nonumber\\
\stackrel{\mathrm{\circ}}{c}_{12}(\eta)&=&(149-887\eta) \;\;\mathrm{GPa}\nonumber\\
\stackrel{\mathrm{\circ}}{c}_{44}(\eta)&=&(105-662\eta) \;\;\mathrm{GPa}
\label{Cstrain}
\end{eqnarray}

and the corresponding hydrostatic strain dependence of the energy-strain coefficients was
\begin{eqnarray}
C_{11}(\eta)&=&(284-1492\eta) \;\;\mathrm{GPa}\nonumber\\
C_{12}(\eta)&=&(149-669\eta) \;\;\mathrm{GPa}\nonumber\\
C_{44}(\eta)&=&(105-761\eta) \;\;\mathrm{GPa}.
\label{Cstrain2}
\end{eqnarray}

The difference between the stress-strain and energy-strain coefficients is quite small, 
and virtually indiscernible on the scale of Fig. \ref{linearC}. Additionally, the use of another exchange-correlation
functional (PW91+VWN, PBE) changed the absolute values of the energy-strain coefficients, but it 
had a negligible effect on the slopes of Fig. \ref{linearC}.

Previously-reported first-principles calculations of the stress-strain dependence of
bcc-Fe (Tab.~\ref{ctbl}) show
a spread of approximately 10\% and tend to overestimate the experimental values for $\stackrel{\mathrm{\circ}}{c}_{11}$ and $\stackrel{\mathrm{\circ}}{c}_{12}$
while underestimating $\stackrel{\mathrm{\circ}}{c}_{44}$. 
This spread may be attributed to \textit{e.g.} different exchange-correlation functionals, applied strains, or criteria for convergence.
We did an interpolation of the previously published data to obtain the values listed for the strain-dependences in the table.
\begin{table}[htb]
\begin{tabular}{|c|c|c|c|}\hline
method&$\stackrel{\mathrm{\circ}}{c}_{11}$(GPa)&$\stackrel{\mathrm{\circ}}{c}_{12}$(GPa)&$\stackrel{\mathrm{\circ}}{c}_{44}$(GPa)\\
\hline
PAW-GGA (present)&284(-1492)&149(-887)&105(-662)\\
PAW-GGA~\cite{nl_th2}&271(-1228)&145(-535)&101(-454)\\
LMTO-GGA~\cite{sha}&303(-1282)&150(-813)&126(-604)\\
PP-GGA~\cite{nl_th1}&289&118&115\\
FP-LAPW~\cite{ma}&285&139&100\\
expt. \cite{rayne}&245&139&122\\
expt.~\cite{jap_adams}&240&136&121 \\
\hline
\end{tabular}
\caption{Comparison of stress-strain parameters (and their hydrostatic strain $\eta$ dependence in parentheses, if available)
obtained in the present study (Eq.~\ref{Cstrain}) with other calculations, and with experimental results.}
\label{ctbl}
\end{table}
Experimental data points~\cite{nl_exp1,nl_exp2,nl_exp3} for the hydrostatic-strain dependencies of the stress-strain coefficients are
displayed along with the calculated results by Sha and Cohen~\cite{sha} in their paper, and these agree well with their 
calculations. Our derivatives do not deviate appreciably compared to the other results.

It is worth noting that the elastic moduli for a wide variety of phases of Fe have been found to 
decrease with applied compressive strain \cite{soderlind}.

\section{Dependence of elastic constants on H-concentration}
\label{sec:Hconcsection}
\subsection{Simulation cells}
\begin{figure}[htb]
\includegraphics[width=6cm]{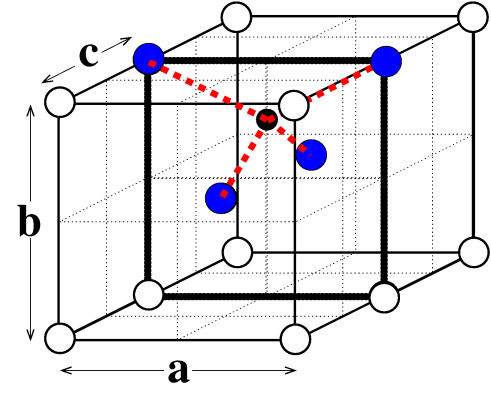}
\caption{Location of an H atom in the tetrahedral interstitial position on the face between two bcc unit cells. Lines towards
nearest-Fe neighbours (blue circles) are shown.}
\label{tetr-fig}
\end{figure}

The elastic constants of pure Fe serve as a starting point for determining the influence of interstitial H atoms on the
elastic properties. 
In this study, we focus on interstitial H in the tetrahedral site (Fig.~\ref{tetr-fig})
because (i) we find this site to be 0.13~eV more stable than
the octahedral one at zero stress (in agreement with previous DFT studies~\cite{JiangCarter,Sanchez}) and (ii) we expect that
the twice as large number of tetrahdedral sites (compared to octahedral sites) per Fe atom will dominate the mechanical
properties at ambient temperatures. 
We implement the variation of H-concentration in the supercell approach of our calculations
by (i) increasing the size of a supercell containing one H atom, or by (ii) adding a second H atom to the
same supercell. In addition, we changed the symmetry of the supercell or the position of the second H atom
relative to the first in order to alter the ordering of the H atoms within the Fe host lattice. The dimensions of supercells 
and number of H atoms used to achieve various H concentrations are listed in Tab.~\ref{supercell}.
\begin{table}[htb]
\begin{tabular}{|c|c|c|}\hline
at. \% H &n(H)&$k\times l \times m$\\
\hline
0&0&1$\times$1$\times$1\\
0.8&1&4$\times$4$\times$4\\
1.8&1&3$\times$3$\times$3\\
2.7&1&3$\times$3$\times$2\\
3.6&2&3$\times$3$\times$3\\
4.0&1&2$\times$2$\times$3\\
5.3&1&3$\times$3$\times$1\\
5.9&1&2$\times$2$\times$2\\
5.9&1&2$\times$4$\times$1\\
7.7&1&1$\times$2$\times$3\\
10.0&2&3$\times$3$\times$1\\
11.1&2&2$\times$2$\times$2\\
\hline
\end{tabular}
\caption{Investigated H-concentrations with corresponding number of H atomsi, n(H), and supercell dimensions
in multiples $k$, $l$, and $m$ of the two-atom bcc Fe cell. For the highest concentration (11.1~at.\%) we
considered three configurations with H-H spacings within the supercell of 4.98~\AA, 3.66~\AA, and 2.31~\AA.}
\label{supercell}
\end{table}
For each of these supercells we determined the elastic constants by
assuming a cubic lattice but, by allowing for internal ionic relaxations, accounting in an approximate way 
for the local distortions due to H.  

\subsection{Effect of non-cubic distortions}
\label{cubicapprox}
The volume expansion of the Fe host lattice by H in the tetrahedral interstitial site introduces distortions
which break the cubic symmetry. For the H-orientation shown in Fig.~\ref{tetr-fig}, the 
Fe nearest-neighbours expand radially-outward from H (along the red lines), resulting in a total expansion,
when projected onto the cube axes, identical in the \emph{a} and \emph{c} directions, and greater than
that along \emph{b} (see \textit{e.g.} Ref. \onlinecite{Beshers}). This tetragonal distortion 
increases the number of unique elastic constants
from three to six: $C_{12}=C_{23}\neq C_{13}$, $C_{11}=C_{33}\neq C_{22}$, and $C_{44}=C_{66}\neq C_{55}$. The variation
among the no-longer equivalent elastic constants depends on the degree of tetragonal distortion, which 
is affected by increasing H-concentration and additionally may be broken depending on the relative H positions in the supercell.

In order to justify our use of a cubic cell, we
compared our results with those from a tetragonal unit cell for one of the most distorted cases considered
in our study, a tetragonal distortion $b/a-1$ of -0.4\%, obtained at an H concentration of 11.1 atomic \%
and H-H spacing within the supercell of 4.98~\AA. Despite the presence
of two H in the simulation cell, the distortion was tetragonal.
%For the cubic unit cells with an H-concentration of greater or equal to 4 at.\%, the 
%elastic constants were explicitly calculated 
%by straining in the different Cartesian directions. For lower concentrations, the spread 
%amongst the three directions was $<$ 2\%
%and the associated errors are from the least-squares
%fitting of the total energy as a function of strain.
The lattice parameters of the tetragonal unit cell were obtained from 
a quadratic fit over a two-dimensional grid of total energies. 
%We calculated the full set of tetragonal 
%or orthorhombic elastic constants by straining the system in two or all three Cartesian directions, depending
%on the number and/or orientation of H in the supercell. 

The elastic constants for the explicitly tetragonally-distorted unit cell at 11.1 at. \%
were calculated using the strain and total energy expressions given in Ref.~\onlinecite{Mehl}. For
the cubic unit-cell, the elastic constants were explicitly calculated
by straining in the different Cartesian directions and averaging.
Table~\ref{tetrtable} summarises the average values of the cubic and tetragonal
elastic constants along with the associated standard errors computed from the spread in the values for the different
orientations. 
\begin{table}[htb]
\begin{tabular}{|c|c|c|}\hline
&cubic&tetragonal\\
\hline
$\bar{C}_{11}=\frac{1}{3}\left(C_{11}+C_{22}+C_{33}\right)$&240$\pm$7&239$\pm$6\\
$\bar{C}_{12}=\frac{1}{3}\left(C_{12}+C_{13}+C_{23}\right)$&145$\pm$3&147$\pm$15\\
$\bar{C}_{44}=\frac{1}{3}\left(C_{44}+C_{55}+C_{66}\right)$&92$\pm$2&93$\pm$2\\
\hline
\end{tabular}
\caption{Elastic parameters (GPa) for the highest-considered H-concentration (11.1 at.\%) as 
obtained for a cubic
or tetragonal (b/a-1=-0.4\%) unit cell.}
\label{tetrtable}
\end{table}
Only $\bar{C}_{12}$ gave a large spread for the tetragonal cell (separately, the values 
were $C_{12}=C_{23}$=138 GPa, $C_{13}$=164 GPa) but not for the cubic cell.
The excellent agreement between the elastic constants of the 
tetragonal and cubic cells is not surprising, as the cell volumes 
were found to be the same, which caused the variations in the elastic constants arising from different 
lattice parameters (details in Sec.~\ref{nonlin}) in the tetragonal cell to mostly average out in the cubic cell. 
As a result of the excellent agreement with the cubic approximation for this highly-distorted case, 
we are confident that our results for all concentrations obtained with the cubic cell accurate. For the 
cubic unit cells with an H-concentration of greater or equal to 4 at.\%, the
elastic constants were explicitly calculated
by straining in the different Cartesian directions and then averaged. For lower concentrations, as the spread
amongst the three directions was $<$ 2\%, there was no averaging done. Instead, the 
associated errors are from the least-squares
fitting of the total energy as a function of strain.

\subsection{Single-crystal elastic constants}

\begin{figure}[htb]
\includegraphics[width=8cm]{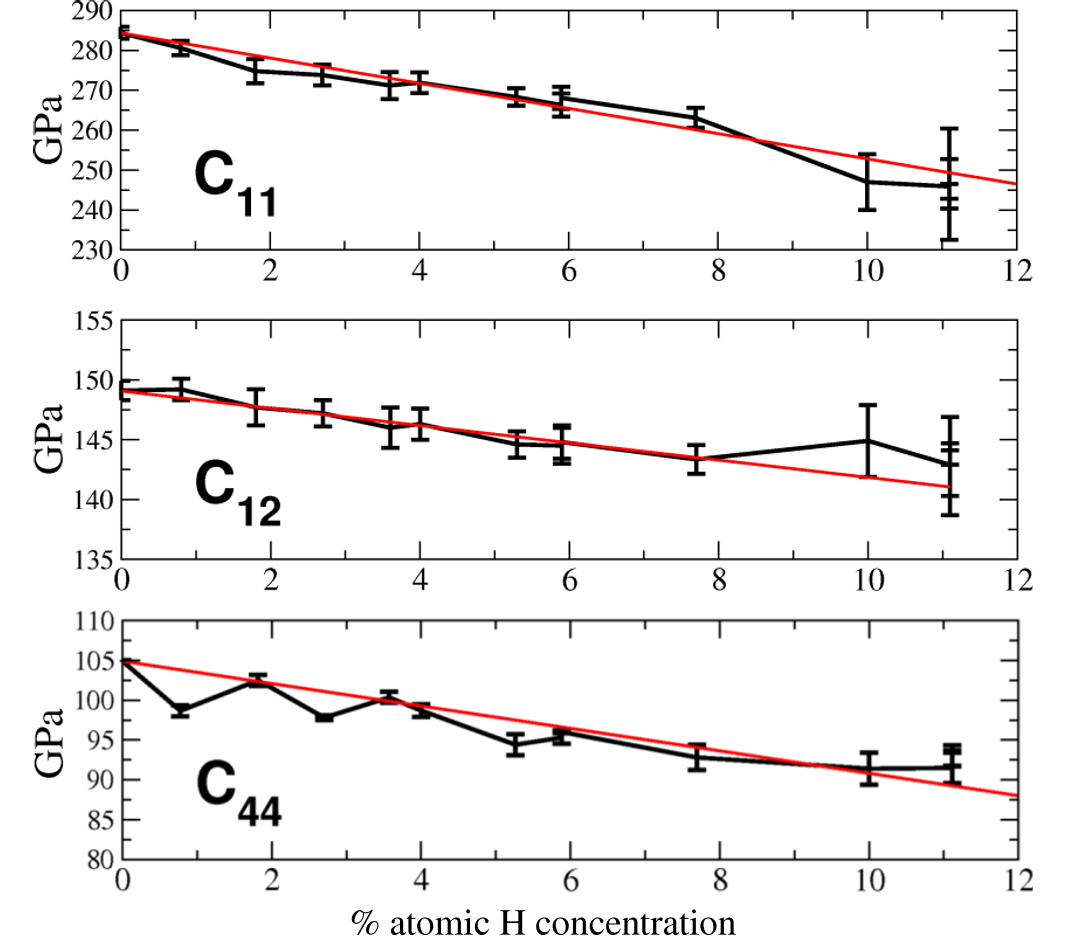}
\caption{Calculated elastic constants for bcc Fe as a function of H concentration, along with
a line of best fit to the $\leq$ 7.7\% points (see text for an explanation regarding the outlying points).}
\label{Cconc}
\end{figure}

The calculated variation of cubic elastic constants with H concentration is shown in Fig.~\ref{Cconc}.
 Additional data points at the same concentration
correspond to differently-ordered structures (see Tab.~\ref{supercell}).
There is a clear trend of decreasing elastic constants with increasing H concentration. 
We ascribe the outlying values of $C_{11}$ and $C_{12}$ for the 10~\% concentration to the high
stress associated with the particular relative orientation of the two H atoms, and also because of their
proximity to their periodic images in the smallest dimension. The spread at 11.1\%
is hidden in the error bars associated with the non-cubic distortions.
At the other concentration, 5.9\%, for which we examined different orderings, 
there was no significant difference between the values. 
We constrained the lines of best fit shown in Fig.~\ref{Cconc} to pass through the zero-concentration value,
but we neglected the data points above 7.7\% and the 0.8\% value of $C_{44}$.
These lines of best fit, as a function of atomic H concentration $x$, are given by
\begin{eqnarray}
C_{11}(x)&=&(284-316x) \;\;\mathrm{GPa}\nonumber\\
C_{12}(x)&=&(149-72x) \;\;\mathrm{GPa}\nonumber\\
C_{44}(x)&=&(105-141x) \;\;\mathrm{GPa}.
\label{CstrainHfit}
\end{eqnarray}

Calculations with an embedded-atom potential of the modification of elastic moduli of Fe by H have been reported recently in
Ref.~\onlinecite{TaketomiCrack}. The data covered up to 6~at.\% H and while the elastic moduli
decreased with H initially, the effect levelled off with increasing H and generally was far weaker than here.

\begin{figure}[htb]
\includegraphics[width=8cm]{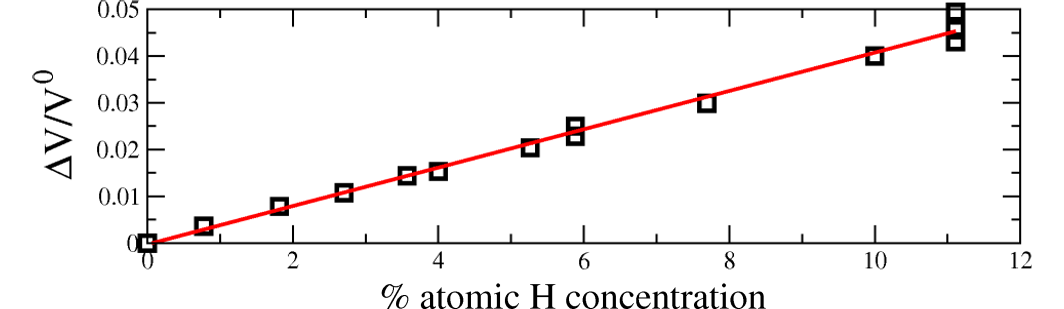}
\caption{Relative volume expansion as a function of H concentration where $\Delta V$ is the
volume difference of the system with H compared with the volume $V^0$ of pure Fe.}
\label{Vconc}
\end{figure}

\begin{figure}%[htp]
\includegraphics[angle=0,width=8cm]{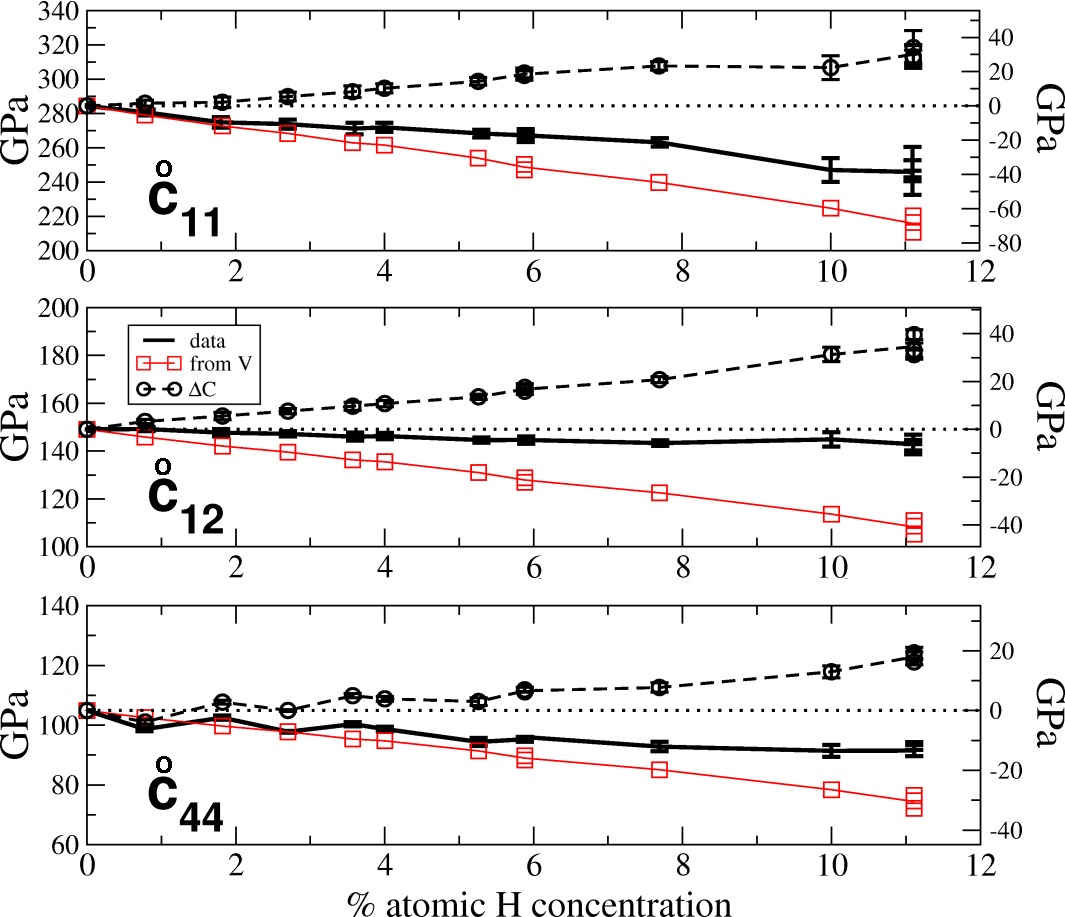}
\caption{Left vertical axis scale: Variation of stress-strain coefficients with H (data from DFT - same as in Fig. \ref{Cconc})
and volume-expected variation (parameterised from pure iron - see Eq.~\ref{Cstrain}).
Right vertical axis scale: difference between total and volume effect (dashed lines).}
\label{concvol}
\end{figure}

Given the pressure-dependence of the elastic constants (\textit{e.g.} Ref. \onlinecite{ma}), one would expect that part
of the observed modification of elastic properties with H concentration can be attributed to the
volumetric effect of the interstitial H on the Fe host lattice. 
In our calculations, the volumetric effect of H is immediately apparent through the linear increase
of supercell volume with H concentration shown in Fig.~\ref{Vconc}. From the slope,
we determined that each H expands the lattice by 4.5~\AA$^3$, in excellent agreement with the value 4.4~\AA$^3$ 
obtained from experiment~\cite{bockris}.
This relation enables us to express the H concentration in terms of a volume change corresponding to
a hydrostatic strain $\eta$ which we can employ in our parametrisation of the elastic constants
(Eq.~\ref{Cstrain}). This correponds to a direct evaluation of the volumetric effect of H on the
elastic constants.  
The volume-induced changes in elastic properties are shown in Fig.~\ref{concvol}, together with
the total effects originally displayed in Fig.~\ref{Cconc}. 
By taking the difference of H-dependence and volume-dependence, 
we remove the contribution of strain of pure Fe from the elastic parameters calculated at the equilibrium
volume corresponding to each concentration of H. 
The resulting difference plots (dashed lines in Fig.~\ref{concvol}) show the residual effects, which include
electronic contributions, of H at each concentration 
(and implicit corresponding volume).  

The separation of solute effects from alloy elastic moduli has been also considered in Ref.~\onlinecite{Buck}
for H in Nb, and by Ref.~\onlinecite{Speich} for different Fe-based binary alloys. In 
these studies, the volume effect was parametrised, by the equivalent
of a line of best fit to the data of Fig.~\ref{Vconc},
whereas we determined it explicitly using the individual data points.

\section{Dependence of strength parameters on H concentration}
\label{sec:moduli}
\subsection{Polycrystalline elastic moduli}
\begin{figure}[htb]
\includegraphics[angle=0,width=8cm]{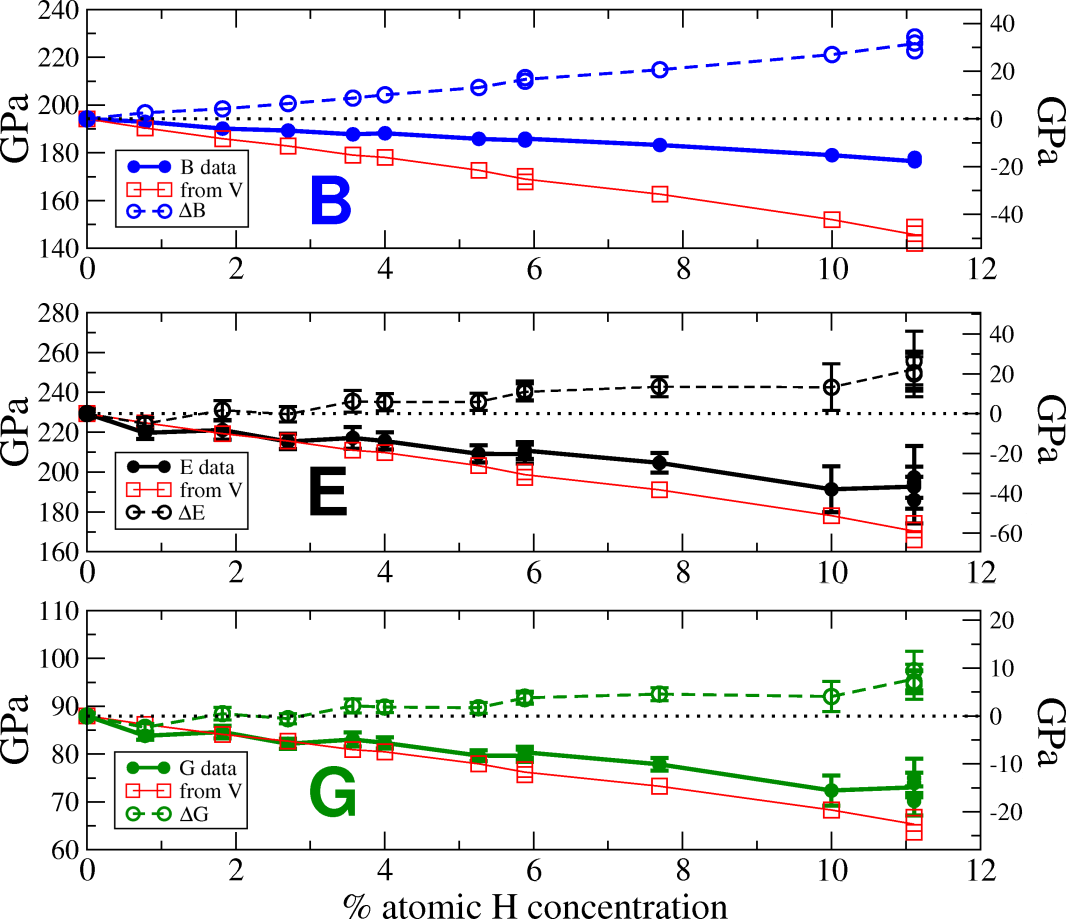}
\caption{Left vertical axis scale: Variation of polycrystalline moduli (B- bulk, E-Young's, G-shear) with H (data from DFT)
and volume-expected variation (parameterised from pure iron - see Eq.~\ref{Cstrain}).
Right vertical axis scale: the difference between total and volume effect (dashed lines).}
\label{polyc}
\end{figure}

Single-crystal Fe samples with H are difficult to prepare while the measurements of the $C_{ij}$
do not directly relate to the strength properties of the material. Most samples are polycrystalline, and
typical measurements are directly related to stiffness (bulk modulus), tensile strength (Young's modulus),
and hardness (shear modulus). In addition, microscopic simulations such as finite-element calculations, whose
inputs consist of polycrystalline averages of elastic moduli, can make direct
comparisons with such experiments.
Therefore, we transformed our single-crystal results to poly-crystalline Fe by using combinations of the
stress-strain coefficients $\stackrel{\mathrm{\circ}}{c}_{ij}$ to derive various elastic moduli 
for describing different types of stress-strain
responses. 
Shown in Fig.~\ref{polyc} are the bulk modulus, and the polycrystalline averages 
of the Young's and shear moduli, where the average was performed according to the expression by Hill,
which is an average of the Voigt and Reuss bounds~\cite{hill}. As in Fig.~\ref{concvol}, the volumetric
changes in the moduli for pure Fe are also shown. 
We find a clear trend in the elastic regime towards a stiffening
(B increased), tensile strengthening (E increased), and hardening (G increased) of the system with increasing
concentration of H. 
Despite the overall softening of the material with increasing H concentration, our findings indicate
that H weakens the softening caused by the accompanying volume expansion.
Our overall findings are in reasonable agreement with the experimentally deduced decrease of the shear modulus
of polycrystalline Fe by 8\% for 1~at\% H~\cite{lunarska}.

The effect of H on the mechanical properties of iron has been controversial. A study by Matsui \textit{et al}\cite{matsui}
found that in tensile tests, the flow stress is increased at low temperatures by H
whereas at higher temperatures it found softening. The study
indicated that the H-dislocation interaction
plays an important role in determining the type of effect that
H has on the mechanical properties. Specifically, whether the presence of H results 
in the hindering or enhancing (as in HELP) of dislocation 
mobility, seems to determine whether the effect is material hardening or softening respectively~\cite{matsui}. 
A very recent discussion of this controversy, which continues to persist, is given
in Ref.~\onlinecite{murakami}. It is important to note that the above-mentioned studies
dealt with austenitic (fcc) steel, and that similar experiments on ferritic (bcc) steel
cannot easily be conducted due to the much greater diffusivity of H in bcc versus fcc iron~\cite{murakami}.

\subsection{Shear moduli in key slip planes}
Our H and volume-dependent elastic constants also enable us to more closely study macroscopic failure 
mechanisms. Therefore, we determine the H dependence of the shear modulus, an important quantity
for describing the stress needed for dislocation nucleation~\cite{Barnoush} and 
glide~\cite{HullBacon}. The shear modulus describes the elastic stress response to applied shear strain.
In a single-crystal sample, the elastic moduli are anisotropic, meaning they take on 
different values when rotated to a different coordinate frame than the standard [100] orientation used
in earlier sections. 

The shear modulus is defined by 
\begin{equation}
G_{ij}\equiv \frac{\sigma_{ij}}{\epsilon_{ij}};\qquad i\neq j;\qquad i,j=1,2,3
\label{shreqn}
\end{equation}
where the indices 1,2,3 denote the $x,y,z$ axes of the coordinate system. 
The transformation of strains from the reference (unprimed)
to rotated (primed) coordinate system can be performed using Euler angles or direction cosines (see \textit{e.g.} Ref.~\onlinecite{HirthLothe}).
The rotated strains hence become $\boldsymbol{\epsilon}^\prime=\mathbf{T}\boldsymbol{\epsilon}\mathbf{T}^\mathrm{T}$ where $\mathbf{T}$ is the transformation matrix for 
transforming the reference coordinates $\mathbf{x}$ into the new frame $\mathbf{x^\prime}=\mathbf{T}\mathbf{x}$. Simplified 
expressions for the directionally-dependent
shear modulus are given in Ref. \onlinecite{TurleySines}, whose notation we follow. 

Using the lines of best fit for the concentration-dependent elastic constants of 
(Fig. \ref{Cconc}), we are able to parametrise the different shear moduli as a function of concentration 
for arbitrary planes and strain directions.
The atomic displacements during a shear distortion constitute slip, 
and in bcc metals, the most common direction of slip is the closest-packed $\langle 111 \rangle$ direction
with the main slip planes being $\left\{110\right\}$ and $\left\{112\right\}$~\cite{HullBacon}.

\begin{figure}[htb]
\includegraphics[angle=0,width=9cm]{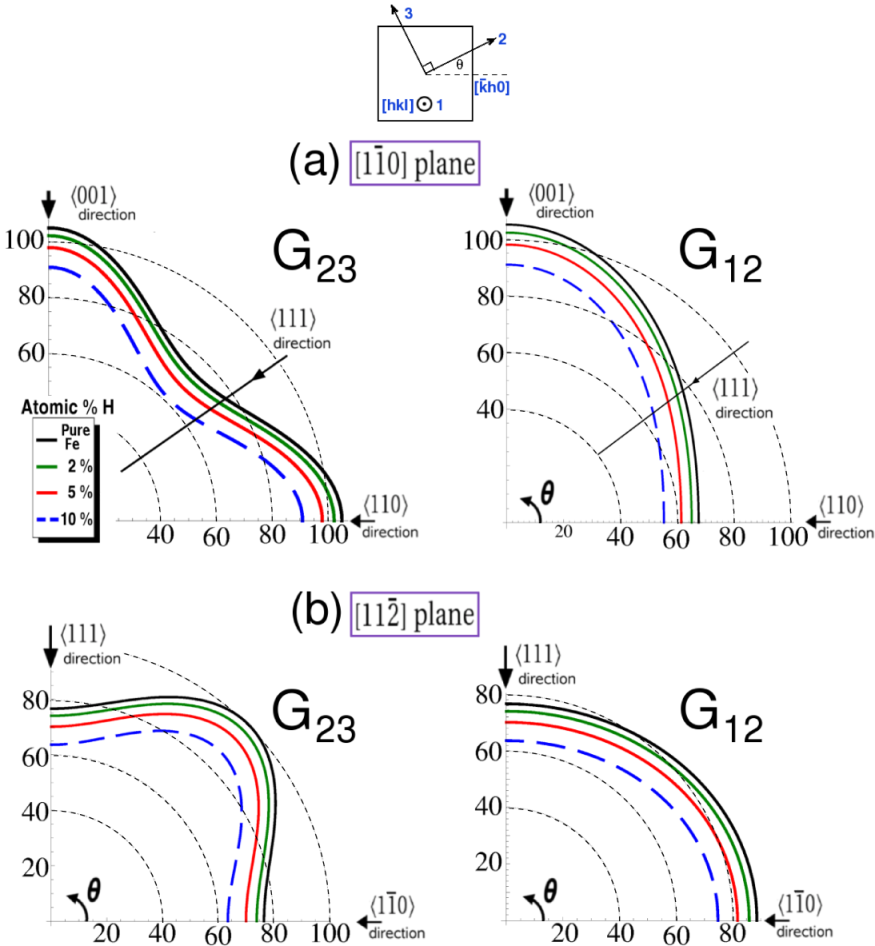}
\caption{Directional dependence of the shear moduli $G_{23}$ and $G_{12}$ (in GPa) for the slip planes
(a) $\left[1\overline{1}0\right]$ and (b) $\left[11\overline{2}\right]$ of bcc-Fe, for different
values of H concentration. The angle $\theta$ is the degree of freedom in the choice of axes in the
shear plane (inset).}
\label{G_23}
\end{figure}

The two major shear moduli for cubic symmetry are $G_{23}$ and $G_{12}$. The 
$G_{23}$ shear modulus describes $y^\prime z^\prime$ shear. The diagram in the upper part of Fig. \ref{G_23} shows the
coordinate system for an arbitrary plane defining the $x^\prime$, or \textbf{1} axis. It contains
a degree of freedom, $\theta$, defining the orientation of the \textbf{2} and \textbf{3}
axes in the plane normal to the \textbf{1} axis. $G_{12}$ is the shear 
modulus when $x^\prime y^\prime$ shear is applied. 

 We display the directionality of the shear moduli as polar plots in Fig. \ref{G_23} for the two key
slip planes of bcc-Fe. The shear modulus $G_{23}$ for pure Fe in the
$\left[1\overline{1}0\right]$ plane
has a maximum for slip in the $\langle 001 \rangle$ and $\langle 110 \rangle$ direction. The minimum, at
$\theta=45^\circ$, does not correspond to any integer-multiple direction; the nearest one being $\langle 111 \rangle$
at $\theta \approx 35^\circ$, which is the expected direction of slip. For the $\left[11\overline{2}\right]$ plane,
the minimum is also the $\langle 111 \rangle$ direction.
The values of $G_{23}$ and
$G_{12}$ are the same for some directions but different in others. They are competing with 
each other when deciding which plane is most susceptible to slip and in which direction. 
For example, looking only at $G_{23}$ of the $\left[11\overline{2}\right]$ plane (Fig. \ref{G_23}b), 
it appears that both the $\langle 111 \rangle$ 
and $\langle 1\overline{1}0 \rangle$ are equally soft, but $\langle 111 \rangle$ prevails in softness when 
$G_{12}$ is examined.
With increasing H-concentration, we find a nearly uniform, linear decrease in the shear modulus in all $\theta$ directions.
The dependence on $\theta$ is weak and not visible on the scale of Fig. \ref{G_23}.
We find a similar rate of decrease for all planes, amounting to an average over $\theta$ of 1.6$\pm$ 0.1\% per atomic \% H.

\section{Conclusions}
\label{sec:conclusions}
We have studied the modification of the elastic properties of bcc Fe by hydrostatic strain and by interstitial
hydrogen. The calculations were carried out for hydrogen concentrations between 0.8~at.\% and 11.1~at.\% with
simulation cells of different dimensions. Our applied constraint of a cubic lattice was verified by a comparison
with a tetragonally-distorted unit cell at the highest investigated hydrogen concentration.

From our density-functional theory calculations, we observe a significant linear decrease of the elastic constants
$C_{11}$, $C_{12}$ and $C_{44}$ with increasing hydrostatic strain or with increasing concentration of interstitial
hydrogen. The volumetric part of the hydrogen dependence can be isolated by relating the volume dependence of the
elastic constants to the volume expansion of the corresponding hydrogen concentration. The overall decrease in elastic 
constants is the result of two opposing contributions: the decrease in elastic constants with 
increasing volume per atom versus an increase from the electronic contribution. These opposing effects may 
help to reconcile contradictory experimental findings - hardening
or softening - under different conditions and concentrations.

We used the elastic constants from our single-crystal \textit{ab-initio} calculations 
to examine the dependence of the polycrystalline elastic moduli $B$, $E$, and $G$ on the hydrogen concentration and find good
agreement with the few available measurements.
The single-crystal shear moduli $G_{12}$ and $G_{23}$ deduced from the ab-initio calculations show an isotropic decrease of
approximately 1.6\% per at.\% H. This suggests that a lower yield stress (assuming the same yield strain) could be
expected as a result of the H-lowered strength parameters in the investigated elastic regime.

\section*{Acknowledgements}
We acknowledge financial support through ThyssenKrupp AG, Bayer MaterialScience AG,
Salzgitter Mannesmann Forschung GmbH, Robert Bosch GmbH, Benteler Stahl/Rohr GmbH,
Bayer Technology Services GmbH and the state of North-Rhine Westphalia as well as
the European Commission in the framework of the ERDF.

\end{document}